# Changes, States, and Events: The Thread from Staticity to Dynamism in the Conceptual Modeling of Systems

Sabah Al-Fedaghi
*sabah.alfedaghi@ku.edu.kw*
Computer Engineering Department, Kuwait University, Kuwait

**Summary**

**This paper examines the concept of *change* in conceptual modeling. Change is inherent in the nature of things and has increasingly become a focus of much interest and investigation. Change can be modeled as a transition between two states of a finite state machine (FSM). This change represents an exploratory starting point in this paper. Accordingly, a sample FSM that models a car's transmission system is re-expressed in terms of a new modeling methodology called thinging machine (TM) modeling. Recasting the car-transmission model involves developing (1) an S model that captures the static aspects, (2) a D model that identifies states, and (3) a B model that specifies the behavior. The analysis progresses as follows.**

- **S represents an atemporal diagrammatic description that embeds underlying compositions (static changes) from which the roots of system behavior can be traced.**
- **S is broken down into multiple subsystems that correspond to *static states* (ordered constitutive components).**
- **Introducing time into static states converts these states into events, and the behavior (B) model is constructed based on the chronology of these events.**

**The analysis shows that FSM *states* are static (atemporal) *changes* that introduce temporal *events* as carriers of behavior. This result enhances the semantics of the concepts of change, states, and events in modeling and shows how to specify a system's behavior through its static description.**

## 1. Introduction

Change is one of the most discussed topics of our time, and scientific journals had published more than a million articles on the topic by the beginning of the 21st century, [1]. Quoting Whitehead, Stickland [2] asserted that change is inherent in the nature of things. Nevertheless, research on change lacks theoretical underpinnings and suffers from an absence of "a process orientation and a wider contextualism."

The lack of these elements essentially reflects myopic and largely unsubstantiated conceptual perspectives on change ([3] as cited in [1]). Numerous works have investigated the concept of change, including reviews across literature regarding change's definition, change relative to time, change conditions and states, the character of the change process, and change in various entities.

The earliest conception of change can be traced to Heraclitus (535 BC–475 BC), for whom all things were in a continuous, ceaseless flux and nothing existed as a static entity. He compared this endless change to a river and remarked, "You can never step in the same river twice" [4]. Other philosophers, such as Parmenides (late sixth or early fifth century BC) and Zeno (495–430 BC), maintained that change is an illusion and that there is just one timeless "being," in contrast to Heraclitus's concept of "becoming" [4]. Several kinds of change have been recognized. For example, Aristotle articulated two kinds of change: accidental change, such as an alteration (e.g., *Socrates becomes pale*), and substantial change (e.g., *the bronze becomes a statue*). Typically, change is viewed as a general notion that is useful in developing ideas that are more specific about change. The concept stimulates critical thinking, which leads to inventiveness and ideas [5].

### 1.1 Change in Computer Science

Computers can be powerful vehicles for change [6]. However, the concept of change is rarely addressed in computer science, except with regard to software engineering and modeling with finite state machines (FSMs). Many domain-specific (computational) ontologies (e.g., OWL) have neglected the notion of change [7]. Change is inevitable for software. Changes in software are implemented to better adapt the software to its environment. In software systems, concerns about change dominate costs at all levels of development as change is adapted to new requirements. As Robbes [8] states, "Systems on which continuous changes are performed inevitably decay, making maintenance harder." This problem is not new: The software research community has been tackling it for more than two decades. However, most approaches have targeted specific maintenance activities using an ad hoc model of software evolution. Robbes [8] proposed "treating change as a first-class entity" through change-based software evolution, in which changes to programs are recorded as they happen.

## 1.2 Change in Modeling

This paper focuses on representing change and its uses in a much more limited domain: conceptual modeling. Conceptual modeling is a central apparatus used in developing systems. In this context, model-based methodologies have been adopted in which a system is represented graphically at several levels of granularity (e.g., UML and SysML). According to Chen et al. [9], a challenge in conceptual modeling is anticipating and accommodating change (e.g., in software or database systems) because "any change of structure, processes and interaction is made through conceptual modeling." Consequently, understanding the concept of change and its related notions (e.g., state, time, and events) is the very crux of modeling.

In pursuing this aim, an entry point into the topic is change in FSM modeling, in which change is viewed as a transition between two states. The most widely used notion in modeling is that of state. FSM is considered a behavioral model, which can be analyzed using a new modeling methodology called thinging machine (TM) modeling to understand change further. Beyond understanding change, another aim of this research is to explore the semantics and expressibility, of both FSM and TM, with regard to related notions such as states and events. Specifically, the focus is on examining how to specify system behavior through its static description. We introduce the concepts of static change and static state, which lead to time-based construction of events.

## 1.3 FSMs as a Conceptual Model

FSMs have been used in software design, especially after the introduction of the extended state machine called a *statechart*, which permits substates of states. Nevertheless, specifying complex state machines can be quite tedious [10]. According to Wagner and Wolstenholme [11], "The concept [of state machine], although born 50 years ago, is still not well understood or interpreted in the software domain, despite its wide application in hardware design. Misunderstandings about state machines have produced several stories and half-truths. The concept of the state machine has been several times (unintentionally?) reinvented for software."

FSMs can be viewed as conceptual tools for modeling a system's behavior as a sequence of transitions, including of time [12]. FSMs are also used to model complex logic in dynamic systems such as automatic transmissions, robotic systems, and mobile phones. Statecharts can represent FSM modeling that allows additional capabilities beyond traditional FSMs such as hierarchical state parallelism [13].

FSMs can change from one state to another, which is called a transition. The concepts of change and state seem highly related; for example, "in change… there is at each stage a moment when the changing item is both in a given state, because it has just reached that state, but also not in that state, because it is not stationary but moving through and beyond that state" [14]. Additionally, FSMs rely on the notions of events, behavior, and time, which are all related to change.

## 1.4 Aim of the Paper

This paper studies and explores the concept of *change* in the context of modeling. FSMs are based on the notion of state, which is very close to that of change. Given that change is missing as an independent concept from conceptual modeling, we use states as a starting point. If FSMs were a type of behavioral model, as is the case with UML and SysML, then further understanding of states would lead to more appreciation of change in modeling.

## 1.5 Outlines of the Approach

Accordingly, state machines are re-expressed in terms of a new modeling methodology called TM modeling. TM modeling is a conceptual tool that abstractly represents a system. It involves capturing (1) static aspects of the system in a model denoted by S, (2) a dynamic representation (denoted as D) that identifies static changes in S, and (3) a behavioral model, B, that specifies the chronology of events.

We provide examples that support TM modeling as a new methodology suitable for all three levels of specification. We can summarize the concepts in this paper in the following steps.

1. A FSM for a car-transmission system (Fig. 1) is selected for analysis.
2. The S TM model for the car-transmission system is presented. Fig. 2 shows a condensed picture of the model, which is shown in full later in this paper.

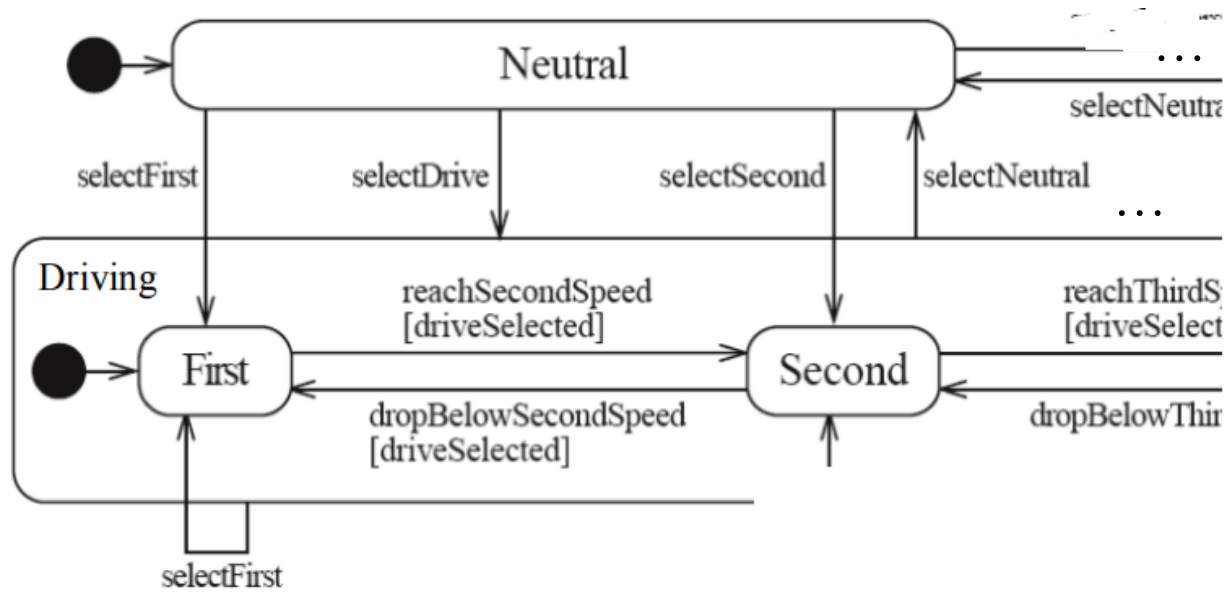

**Fig. 1. State machine of a car transmission (partially from [15]).**

3. Fig. 3 shows a similar picture of the D model of S.
4. The behavioral model B is extracted from D.

The crucial analysis step of this multilevel modeling involves the move from staticity in S (Fig. 2) to staticity in subsystems, which are shown as colored subdiagrams in D (Fig. 3). The general transformation of this process is as follows:

Static S
↓
Static **changes** (multiplicity in S with no order)
↓
Static **states** (static changes with order, D)
↓
**Events** (static states with time order)
↓
B

The main observations in this process are as follows:

- S only represents the steady (static) whole, so it is necessary to analyze the underlying decompositions where behavior can happen (potentiality of dynamism).
- Dividing S causes multiple subsystems to be created (see Fig. 3), which convert the system in *static changes* (constitutive components) with respect to the whole S.
- For static changes (the colored areas in Fig. 3), multiplicity is a *form of becoming* from the unity (S). A static change here is analogous to a set that is replaced by its members.
- Static states are ordered static changes.
- The states in FSM modeling are types of static states; thus, they and their transitions *cannot represent the system's behavior* (i.e., they do not introduce time).
- A system's behavior is specified by introducing time into its static states, thus converting them to events and fixing their chronology, producing the B model.

The aim of this discussion is to understand what a system state is, what is involved in the notion of change, and how change is related to states. Furthermore, we seek to understand how a static description is made of a dynamic specification of a system behavior, how to create multiple subsystems from a single system, and the roles of time and order in this arrangement.

The enhanced review in the next section summarizes the general features of the TM model, which is a promising modeling approach that can be applied in diverse applications such as designing unmanned aerial vehicles [16], documenting computer networks [17], modeling network architectures [18], modeling advanced persistent threats [19], modeling an IP phone communication system [20], and programming [21].

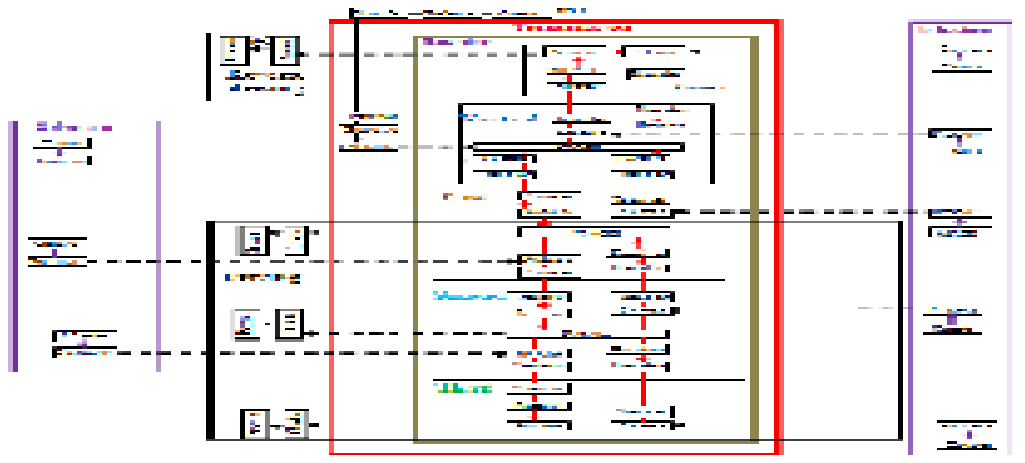

**Fig. 2. A contracted view of the schematic S model that corresponds to a state machine of a car transmission in Fig. 1. The full details of S will be shown later in this paper.**

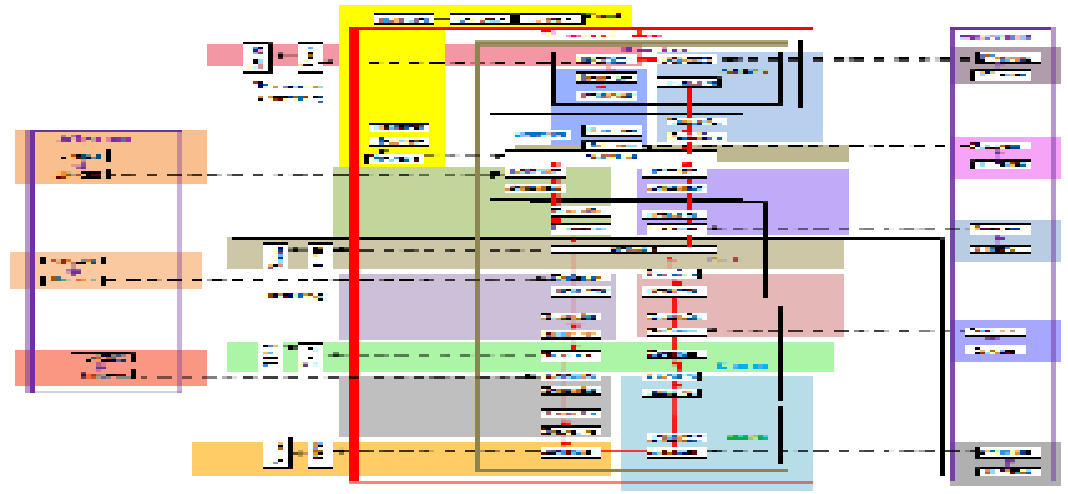

**Fig. 3. A contracted view of the D model that corresponds to S in Fig. 2.**

The TM model can also be used to model service-oriented systems [22], business systems [23], a tendering system [24], robotic architectural structure [25], the VLSI engineering process [26], physical security [27], the privacy of the processing cycle of bank checks [28], a small company process [29], wastewater treatment controls [30], asset-management systems [31], IT processes using Microsoft Orchestrator [32], digital circuits [33], and automobile tracking systems [34]. The remaining sections discuss how the TM model can be applied in analyzing changes and FSMs.

## 2. Thinging Machine Model

Imagine that, for various reasons, you are not satisfied with the ontology used in object-oriented modeling (e.g., [35]). Before describing or embracing a new model, you must adapt a certain conceptualization of the "domain view" (relevant to a particular sphere of interest, e.g., accounting or tourism). Specifically, in the context of this model, you must determine *what your ontology is*. That is, what are the things in your model, what is their order, how are they constructed, and what basic presuppositions underlie the model? Note that, by contrast, philosophical ontology starts with what *exists*.

Ontologies are frameworks used to represent shareable information and knowledge (e.g., Semantic Web [36]). An ontology is a "specification of a conceptualization" within the involved domain [37]. In modeling, ontologies give descriptions of the "things that are modeled" in a particular domain area. However, an ontology not only is a set of categories but also involves a structure that includes relationships (e.g., is–a), order (e.g., part–of), and constraints.

An ontology matters as a workable means of communicating, planning, and guiding the development of projects (e.g., building and controlling a system such as the car-transmission system modeled later in this paper). We model such a system using TM modeling, which starts with declaring the kinds of things used in a TM. In contrast to object-oriented modeling, TM modeling does not include notions such as objects, properties, or relationships.

Deciding the sorts of things in a model involves understanding the *categorical* structure of the modeled domain under consideration, which is typically hierarchical (e.g., classification). The *number of categories* may distinguish different ontologies. Aristotle (384–322 BC) articulated *10* categories of things in the world. An example of the *three-category* ontology is the state machine categories (states, events, and transactions). In philosophy, a *one-category* ontology exists that only includes so-called *tropes*. According to Paul [38], "One category ontologies are deeply appealing, because their ontological simplicity gives them an unmatched elegance and sparseness.… We don't need a fundamental categorical division between particulars, individuals, or space-time regions and their properties, nor do we need a fundamental categorical division between things, individuals, or bearers and the qualities 'borne' by them."

TM modeling has one category called *thi*ngs/*ma*chines (*thimacs*). Note that this study offers an idealization and, sometimes, rational and linguistic arguments (no data, as in physics research) closely reflecting philosophical-like (computer science philosophy) speculations. The proposed ontology in terms of thimacs is a deliberate simplification of a modeled domain's description used to identify core concepts in modeling.

One of the main interests of this paper is understanding change in the context of the TM thimacs. Specifically, we focus on the change in the thimacs in the S model when time is introduced to convert them into event thimacs.

## 2.1 Basic TM Model Constructs

A thimac in TM modeling is denoted as Δ, which has a dual mode of being: the machine side, denoted as M (see Fig. 4), and the *thing* side, denoted as T. Thus, Δ = (M, T). The S model is the grand thimac, with a subthimac structure comparable to classes and subclasses in object-oriented modeling.

The notion of T relies more on Heidegger's [39] notion of "things" than it does on the notion of *objects*, with the latter being a very popular notion in computer science (e.g., object-oriented modeling). M refers to a special abstract TM (Fig. 4), which exists as a basic, complete machine. The thimac Δ is *the TM element of modeling* that reflects an object/process and a product/productivity simultaneously. Conceptually, such a picture implies two facets: a *being* (in the context of the model) as the thing and its passage (machine) to being (thing). The thing and the machine are like the faces of a faceted jewel, in that the thimac retains its unity simultaneously embracing a plurality of facets. For example, water is a water-thing, and its machine is its processual configuration (organization), which involves oxygen and hydrogen and leads to its manifestation. The machine is written as $H_2O$ in shorthand, which indicates a process that generates a unity.

Philosophically, the thimac is "being/becoming." According to Zubiri [40], the process "would be the *inner, intrinsic joining* of what we call 'being' and what we call 'non-being'" (italics added). We interpret this joining from nonbeing as a passage from one condition to another.

A machine can be a subdiagram of the diagram of Fig. 4 (e.g., it can only create and process things), or it can be a complex of these machines. M is built under the postulation that it performs five generic actions (creating, processing [altering], releasing, transferring, and receiving) or a subset or complex of these actions. A thing is created, processed, released, transferred, and/or received, whereas a machine creates, processes, releases, transfers, and/or receives things.

The five actions (also called stages) in Fig. 4 form the foundation for Δ-based modeling. Among the five stages, flow (a solid arrow in Fig. 4) signifies conceptual movement from one machine to another or among the machine's stages. The stages can be described as follows.

- *Arrival*: A thing reaches a new machine.
- *Acceptance*: A thing is permitted to enter the machine. If arriving things are always accepted, then arrival and acceptance can be combined into a "receiving" stage. For simplicity, this paper's examples assume that a receive stage exists.

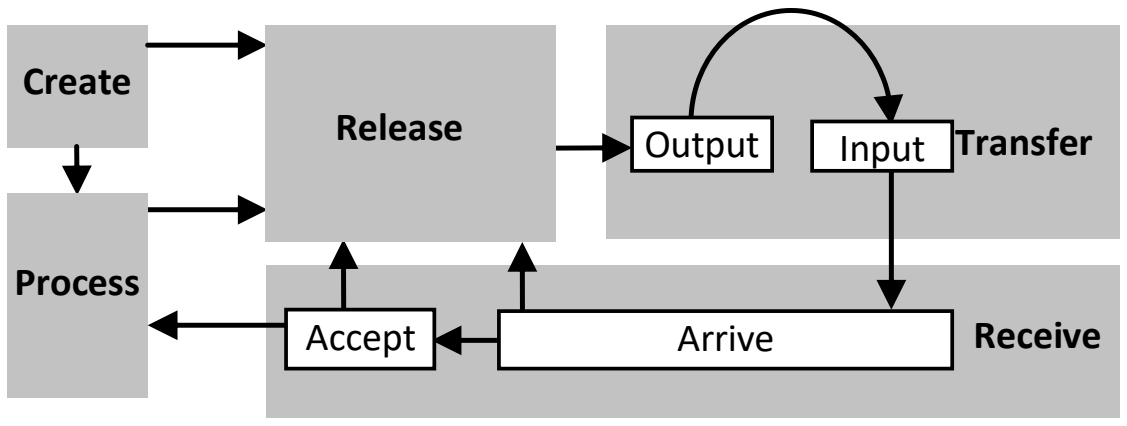

**Fig. 4. The thinging machine M.**

- *Processing* (alteration): A thing undergoes modifications without creating a new thing.
- *Release*: A thing is marked as ready to be transferred outside of the machine.
- *Transference*: A thing is input or output outside of or within the machine.
- *Creation*: A new thing is born (created) within a machine. Creation can designate bringing into existence (e.g., ∃ in logic) in the system because what exists is what is found. Creation in M indicates "*there is*" in the system but not at any particular time.

The machine T can be simplified as shown in Fig. 5.

The TM model also includes the notion of *triggering*, which connects two subdiagrams where there is no flow between them. The triggering is represented by dashed arrows in the TM diagram. The TM model can be specified in a textual language, wherein the arrows are represented by dots. For example, the different flows in Fig. 4 can be specified as follows:
*Flow.create.release.transfer.output*
*Flow.create.process.release.transfer.output*
*Flow.transfer.input.receive.arrive.release.transfer.output*
*Flow.transfer.input.receive.arrive.accept.release.transfer.output*
*Flow.transfer.input.receive.arrive.accept.process.release.transfer.output*

### 2.2 TM Example

According to Busse et al. [41], the Aristotelian categories were accepted for quite a long time. Other categories were only introduced as subcategories. In the 19th century, the additional category "facts" was added. For example, *A phoned B on May 23, 2012, at 2:11pm* is a fact [41]. In TM ontology, it is an event. Fig. 6 shows the TM model S for *A phoned B*. When time is considered, the event *A phoned B on May 23rd, 2012, at 2:11 pm* occurs as shown in Fig. 7. Figs. 6 and 7 show the machine side of the thimac *A phoned B* and its corresponding time subthimac.

In this paper, we investigate the transformation shown from Fig. 6 to Fig. 7 to better understand how to arrive at a dynamic specification of a system. In such an examination, the notion of (static) changes and (static) states appears during this transformation to *A phoned B on May 23, 2012, at 2:11 pm*.

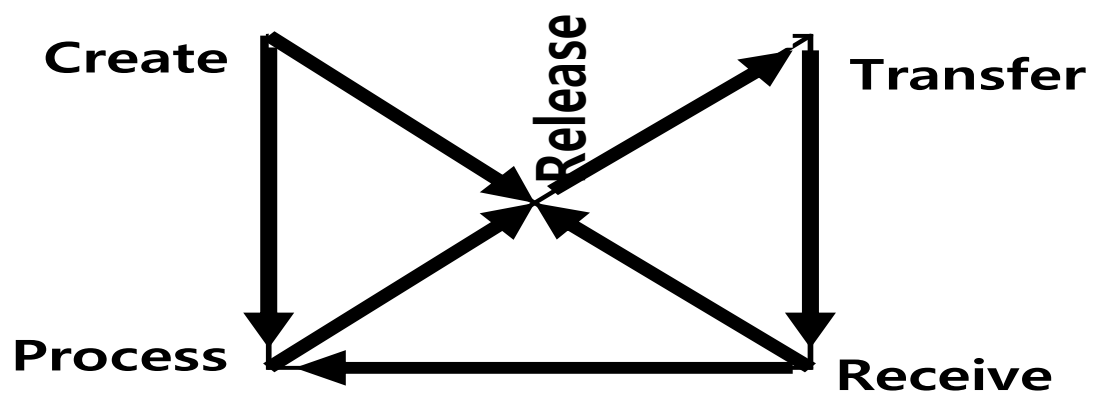

**Fig. 5. Simplification of M.**

## 3. First Phase of Modeling: TM Static Model

Badreldin [15] introduced a state machine example (Fig. 1) of a car-transmission system comprising a two-level nested state machine. The transmission starts in the neutral state. While in the neutral state, the state machine responds to four FSM events, namely the selectFirst, selectDrive, selectSecond, and selectReverse events. Each FSM event triggers a transition to a new FSM state. While in the Second state, the transmission system responds to two events—reachThirdSpeed and dropBelowSecondSpeed—which trigger transitions to the ThirdGear and FirstGear states, respectively [15].

Fig. 8 shows the static TM model S of the transmission system. S "pops up" into existence from the modeler's conceptualization piece by piece (it may not be written in language at first) as the modeler looks at the patterns dance before their imagination. In TM, these patterns come in terms of thimacs that are expressed as in S.

This process of thinking in terms of thimacs begins with the car starting (1), which creates a signal that reaches the transmission (2) that is processed (3) to trigger the neutral process state (4). At this moment, the transmission is ready to move from the neutral position.

- In the neutral position, upon the driver selecting the first position (5), the transmission (gear) moves to the first position (6 and 7).
- The processing of the first position (8) triggers (dashed arrow) the first driving condition (Driving label in the diagram) for the car (9).
- Upon the driver selecting the second position (10), the transmission (gear) moves to the first position (11 and 12).

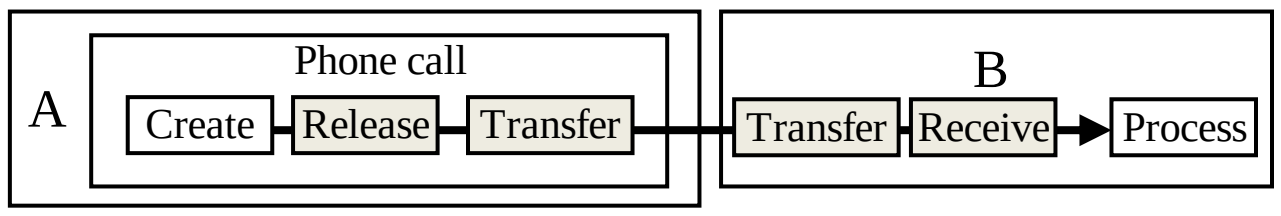

**Fig. 6. The TM representation (S model) of A phoned B.**

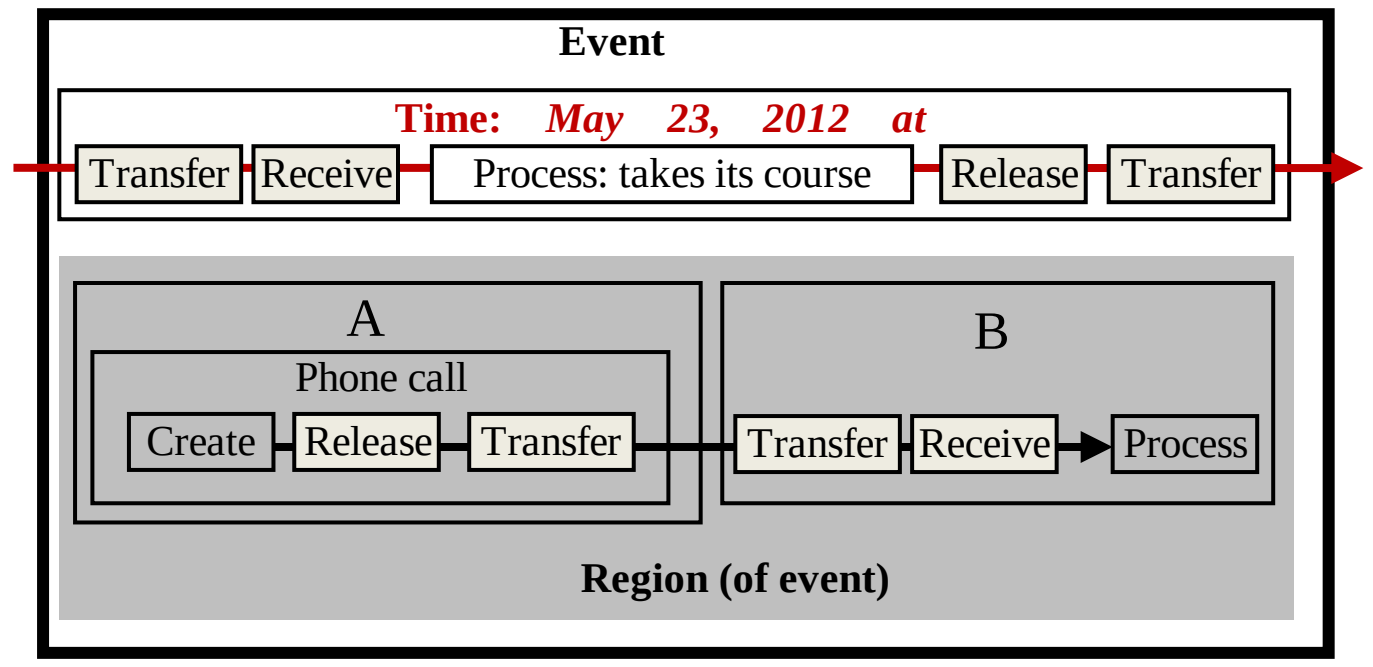

**Fig. 7. The TM event *A phoned B on May 23, 2012, at 2:11 pm*.**

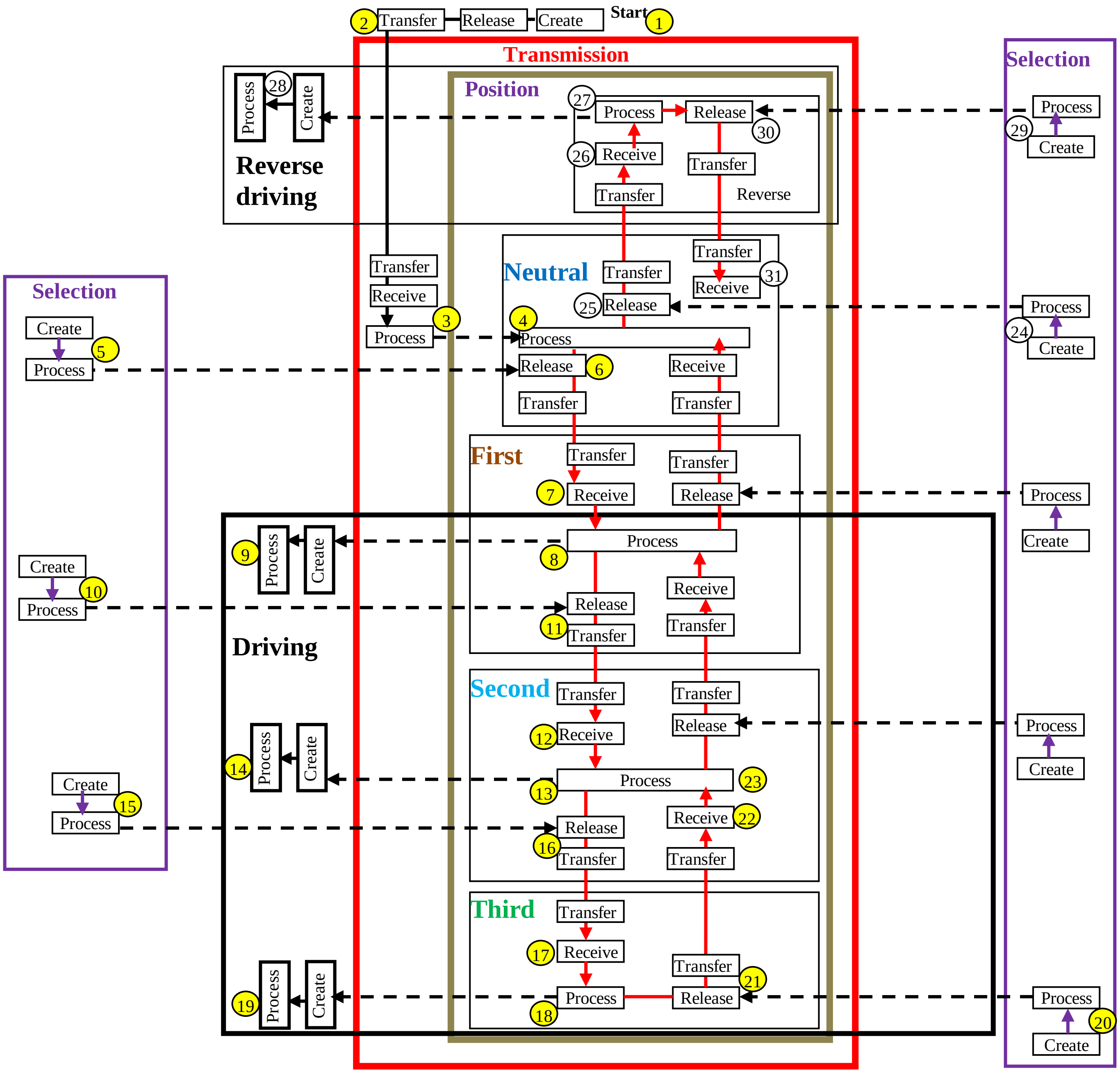

**Fig. 8. The TM representation (S model) of a car's transmission system.**

- The processing of the first position (13) triggers the second driving condition for the car (14).
- Upon selecting the third position (15), the transmission (gear) moves to the third position (16 and 17).
- The processing of the third position (18) triggers the third driving condition for the car (19).

Similar descriptions are applied when moving the position from third to neutral. To avoid repetition, we only describe one such action: from the third position to the second.

Upon the driver selecting the second position (20), the transmission (gear) moves to the second position (21 and 22).

- The processing of the second position (23) triggers the second driving condition for the car (14).

Back in the neutral position, the transmission is ready to move to the reverse position.

- Upon the driver selecting the reverse position (24), the transmission (gear) moves to the reverse position (25 and 26).
- The processing of the second position (27) triggers the reverse driving condition for the car (28).
- Upon the driver selecting the neutral position (29), the transmission (gear) moves to the reverse position (30 and 31).

## 4. Steps to Dynamism: Decomposing S

Behavioral states and transitions appear in Badreldin's [15] state diagram without any explanation for how they emerged from the English description of the transmission device. Do the states in Fig. 1 represent parts (fragments) of the transmission system? If they do, then how are these *static* parts converted to *temporal* states that represent the system's behavior? In this discussion, we try to develop a theory for transformation in the TM model, from a static description such as S to a description of dynamic behavior. Accordingly, in this section, we analyze the roots of dynamism in static modeling (e.g., S). The results of this analysis are applied to model the transmission system's behavior.

S (Fig. 8) is a static description that represents a stillness or rest (no time) condition. It lacks a type of structure that applies "meaningfulness" (which is explained later) to its parts. This S/parts requirement is reminiscent of Deleuze's philosophical notion of a "body without organs," but in our case (in contrast to Deleuze's aim of *immanence*), we aim to identify "static organs" (parts) to specify a dynamic system of S. The static organs facilitate the roles of the release, transfer, and receive stages in the TM model (i.e., communication with other machines). The body without organs is a thimac with only the create and process actions.

From a different perspective, S is a machine schema that is amenable to compositional exploration to generate a new structural level (multiplicity). Building on Maturana and Varela's [42] work, the *organization* notion defines a system (e.g., S) as a unity that outlines its form and serves as its core identity. According to Whitaker [43] (describing Maturana and Varela's [42] ideas), "A system's organization defines its identity, its properties as a unity, and the frame within which it must be addressed as a unary whole." On the other hand, the *structure* of a particular composite unity is the manner in which it is made by actual static components in a particular space as well as a particular composite unity (this description is a modified version of the notion of structure in Maturana and Varela [42]). The point of this discussion is to view S as an organization that needs structure so that its behavior can be specified. While the wholeness of S is the same, S may have different structures, depending on how it is divided into parts.

The concept of whole-multiplicity helps in forming an assemblage of the fragments that evolve to facilitate dynamism. Dynamism, in this context, refers to ordered temporal events. To partition S, we create multiplicity through parts that emerged from the whole. This emergence may be conceived of as a qualitative change that consists of the appearance of "things of a new kind or ontological species" [42]. The selected subdiagrams of S become new thimacs, and the original thimac becomes a network of subthimacs. This is an evolutionary change from thimac S to its parts/subthimacs utilized to identify the abstract notion of the system's behavior. Evolution refers to change simpliciter [44].

S is a complex entity consisting of many subthimacs interconnected in some specific way. The system's characterization (both static and dynamic) resides not only in the separate subthimacs but also in the structure they form. The new parts form a new structure. In the state machine phase of design, the states of the transmission machine are identified (by the modeler, in a vague way) and connected together to characterize the transmission behavior. In the example, the transmission thimac consists of the subthimacs first, second, third, neutral, and reverse. The first three states form a subthimac called driving.

S becoming (forming) multiple parts with different properties is analogous to a tree in the fall, which experiences many changes occurring simultaneously: red leaves, yellow leaves, dry leaves, etc. In the transmission example, the transmission machine is transformed into many submachines: the FirstGear machine, SecondGear machine, ThirdGear machine, NeutralGear machine, etc. Accordingly, a transformation occurs from sameness to different parts. This is not an easy task because no clear borders mark where the cutting occurs. However, this is better than the FSM method because we have a diagrammatic description instead of simply text.

### 4.1 Justification for Decomposition

(a) Decomposition is necessary because the system described by S is obviously "provoked" behaviorally, piece by piece (subdiagrams); for example, in a FSM, a state at a time (at this point, we ignore the order of the S parts being activated). TM modeling produces a single, whole description of the system to avoid any inconsistency; hence, after this whole is generated, the description is divided into pieces to define the dynamism of its interiority. The division is performed to pursue a (correct) variety that produces a blend of staticity and changeability. While staticity is conserved, changeability emerges from differences of parts, just as cutting an ice block into smaller blocks preserves coldness. Importantly, the gestalt of the parts must be assembled into the original totality of the system during the dividing process. The original single machine of the transmission system is replaced by an assembled machine that is formed from selected submachines.

(b) The overall work of the transmission system is assigned to distinct subthimacs (subsystems); therefore, understanding the system requires identifying these thimacs and their contributions to this overall work. In other words, the system description S needs to be regenerated in joinable parts that come into reflective relation with the whole. This causes the system to be restructured under the influence of parts.

(c) Exposing the order is important because S represents only the whole, so the underlying decomposition must be revealed where behavior potentialities may happen. The selected parts may have originally been thimacs, several thimacs, portions of thimacs, or a mix of thimacs and portions of thimacs. These parts of S allow one to understand the system's behavior, which involves activating certain prearranged subdiagrams, whose semantics are determined by how the subdiagrams are interconnected. Importantly, the decomposition of S exposes orders (i.e., before, after, and simultaneous) among parts of S that are recognizable by the flows and triggering among the parts (subdiagrams). These relations make a difference when specifying the system-level behavior. Each part of the system has a distinct role, and the parts are interdependent based on their position in S. Our aim is to identify these interdependencies. S is an atemporal model, so the interdependencies are static relations among parts of S.

(d) Directional control: Decomposing S into parts causes a multiplicity of subsystems to be created that convert the system into constitutive components with respect to the whole, to anticipate and infer what is expected. In the transmission example, suppose that the car is idling and the neutral state is active. Then, the system control concentrates its processing activity on a "shift to one" or to "reverse" and neglects all other parts (two and three) that have no direct bearing upon the expected next state. When the transmission is shifted to the "reverse part," the focus will be shifted to "driving in reverse" or "moving back to neutral." The other parts are ignored. Note that parts of S such as FirstGear and SecondGear, etc. are representations that are taken as a base for dynamic specification of a concrete phenomenon in a particular instant of time in the D and B models.

## 4.2 The Change: Whole to Multiplicity

A *change* ordinarily refers to two meanings:
(a) An event of changing (typically called a process); for example, the elevator door is opening.
(b) A *state* of being a change (e.g., the door is open).
These types of change involve the "same thing becoming" (Aristotle's words) different. Aristotle mentioned another type of change: *becoming from nonbeing to being*. In this change, *nonbeing* is a different thing from the *being*. Similarly, the unity (S) becomes multiplicity as a result of multiple becoming of S into its parts (we call them *static changes*). Change here means *variation* in the sense of different parts of S (e.g., *changing* a bill into coins).

Fig. 9 shows a selected division of S for the transmission system into 22 static changes. This division produces distinct parts of S. An analogy is a die that has six changes corresponding to its six faces. The dynamism of the die originates from *conceptually* dividing it as a whole cube and replacing it by its six faces. Similarly, the aim of dividing S is replacing it with a certain number of changes. Yet, a better analogy is a chess game, which features many changes. A change in this case is a *part* of the chessboard's setup or its static description (S of the chess). Accordingly, when specifying the chess tree (as in artificial intelligence) of the chronology of legal changes allowed in the game, the tree is not related to the behavior but to the static arrangements of parts relative to each other. In another example involving numbers, the sequence *2, 3, and 4* expresses a logical relationship, not a temporal one.

The point of these analogies is to project the division of the whole (e.g., a die) onto dividing S as a static description that embeds its possible changes. Identifying these changes and their relations produces an atemporal description. Hence, identifying states (types of changes) in Badreldin's [15] state diagram does not involve the system's dynamism (temporal-based order) or behavior.

A static change in the S model is not the common notion of so-called dynamic change as the *process* of causing a thing to become different from what it is at present or what it was in the past. This process change refers to *becoming* different or *becoming* altered or modified. In the TM multiplicity paradigm, static change refers to the feature of *being* different. In the static atemporal progressions in S, there is no becoming (a process) but only being (a thing). Dividing S into a form of multiplicity produces atemporal changes with no specific order (e.g., when using a multicutter to cut an apple into pieces, all of the pieces are produced simultaneously).

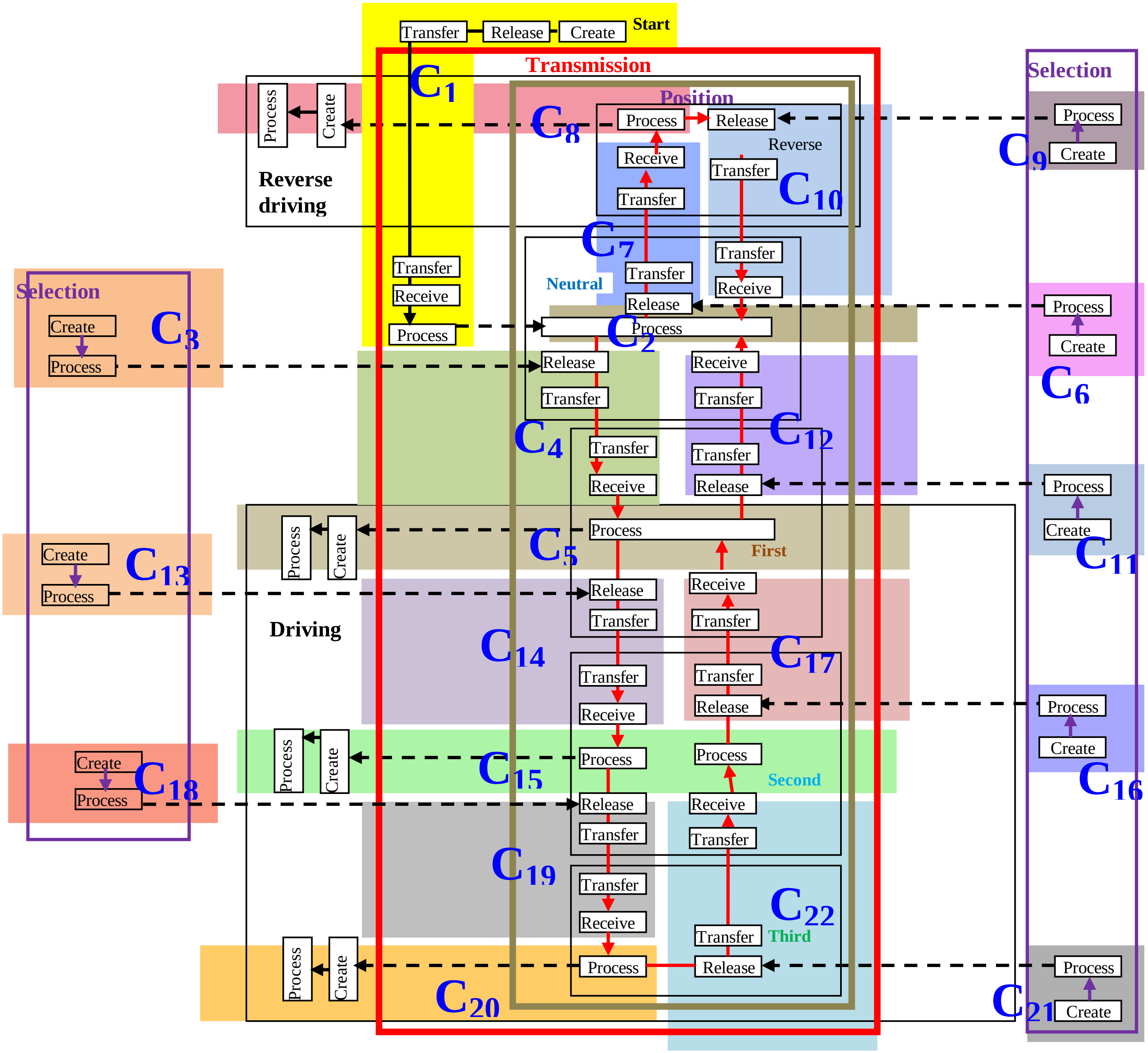

**Fig. 9. The TM representation (S model) of the transmission system.**

## 4.3. Required Properties of Static Changes

The division should produce a juxtaposition (a topology of parts) with parts that are sufficiently "meaningful." The meaningfulness of a part of S resides in the isomorphism between the part and the thing it is supposed to represent in reality (in the modeler's conceptual framework). For example, in the context of changing the transmission from one position to another, "release" by itself as a subdiagram does not seem to have this meaningfulness. "Release and transfer" seems to be a more meaningful part, but "release, transfer, transfer, and receive" is an ideal whole/part because it corresponds to the familiar notion of "moving from… to…," as in the transmission moving from the NeutralGear to the ReverseGear positions. This *moving from… to…* part of S is an example of the (partial) "whole" that we are looking for in dividing S. Note that it is formulated from two halves of different thimacs. In general, the initial "elements" (e.g., thimacts) of the whole (S) are not the best 'carving' suitable for our purpose.

When modeling the state diagram of the transmission system, Badreldin 15] unconsciously followed a "meaningful" criterion. Badreldin's [15] state diagram seems to be a coarse-grained description because the classical definition of a state and the mixing up of states and transitions create a confused vision that hides the granularity of static description. Ontologically, the state diagram is a "states and transitions diagram" in which transitions are labels on the diagram's edges. This distinction between states and transitions is an outcome of the separation of object/process in the object-oriented paradigm.

In TM modeling, states and transitions are thimacs. In the scheme to divide S, operators do not need to transform one state into another because the operators are parts of S, and connections among parts (states) are pre-specified by flows (inputs and outputs) and triggering. Thus, as is shown later, the behavior model D has no labels on the edges. The multiplicity of S encompasses states, transitions, and other parts as one category: subthimacs (subdiagrams) of S. Fig. 10 shows the conversion of changes to TM states and the order imposed on these states.

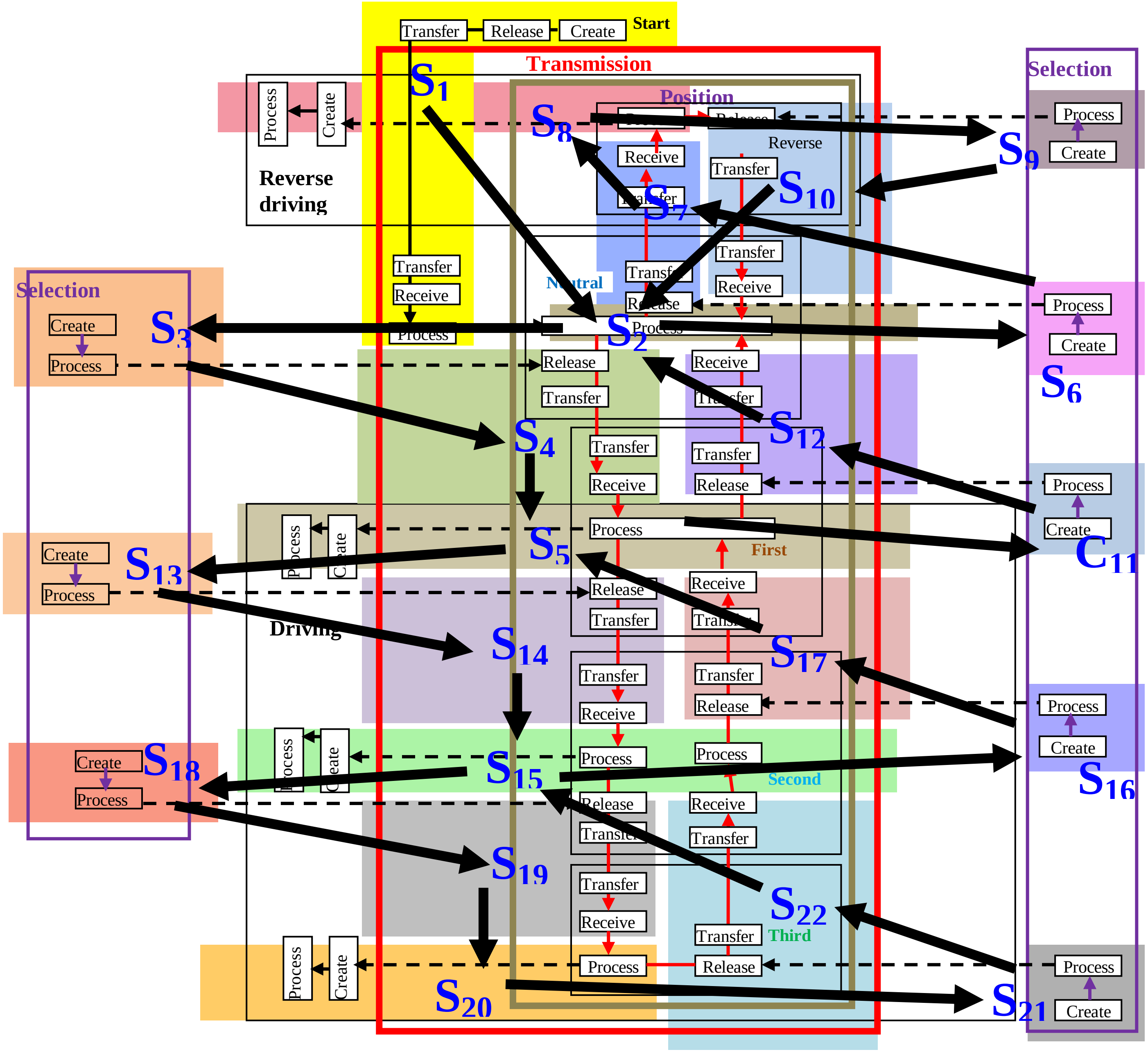

**Fig. 10. The TM representation (D model) of the transmission system.**

## 5. TM States: Ordered Static Changes

The aim of this discussion is to understand how to identify a system's *states*. However, the static change notion under consideration is more general than the classical notion of a state. The basic idea is that a system is divided to allow anticipation of its behavior. The supposition is that multiplicity can be represented as a chronology of states.

Fig. 11 shows the chronology of the transmission system's states.

Static changes and static states are different, in terms of their context of simultaneity and order. Whereas static changes are created simultaneously, states are changes with an atemporal order. Thus, we can develop a chronology of states but not of changes. Note that the FSM distinction between a state and a transition is not applied in TM modeling. The transitions, as well as FSM states, are static changes. For example, in Fig. 7 (the transmission model), the transition from first gear to second gear is represented by the subdiagram *One.release.transfer.Two.transter.receive*; thus, it is one part (static change) of S. This subdiagram is a static state when order is imposed on all static changes. In FSM modeling, only one FSM state is active at any given time. Ambiguity exists in the meaning of this transition with regard to the object's condition between two FSM states. The transition seems to have an existence comparable to that of states.

Consider the transition between the liquid and vapor states of water. Boiling takes time and in many physics texts, it is called the boiling *state*. The boiling water stays partially liquid and partially vapor. This is similar to what is called *entanglement* between systems in quantum theory, as demonstrated in the famous Schrödinger's cat puzzle. In the TM model's rough macroscopic static states, the *observing* of the entangled mixed state is considered a static state because the whole scenario is timeless. Hence, the subdiagram described by *One.release.transfer.Two.transter.receive* is considered a state in its order within static states.

Fig. 10 specifies static states based on changes. Consider the following two changes, assuming the car has started:
C′ : *The transmission* is moved *to the first position*.
C″ : *The transmission* is moved *to the reverse position*.
A subdiagram of Fig. 10 represents each of these changes. In terms of "after," "before," and "simultaneously," it is unclear how to order these changes. In contrast, when given *7, 8, and 9* (or any three different integers), we can establish the ordering (e.g., in ascending or descending order). If we replace C″ with C‴: *The transmission is moved to the third position*, then it is easy to observe that C′ is before C‴. Note that the relationship is atemporal (similar to relationships between numbers), whereas a point, a line, and a square embed some atemporal ordering.

## 6. Events: It Is Time to Introduce Time

Changes and TM states form the foundation upon which to understand events. An event is a period when a thimac materializes. We have projected the thimac materialization in terms of its subthimac's (changes/states) materialization. Car travel is a thimac, and a car traveling at a particular time is an event (a thimac with a time subthimac). Car travel has many subthimacs; hence, these subthimacs have many "small" events that comprise the car travel event, including the car's transmission events. To identify these small events, the transmission is replaced by its parts (static changes), which are infused with an order to produce small, correlated static states. Lastly, these states are altered by inducing time to be transformed into events. A phenomenon (e.g., acts that happen to the car/by the transmission) undergoes transformation from staticity to dynamism in terms of events, as illustrated in Fig. 12.

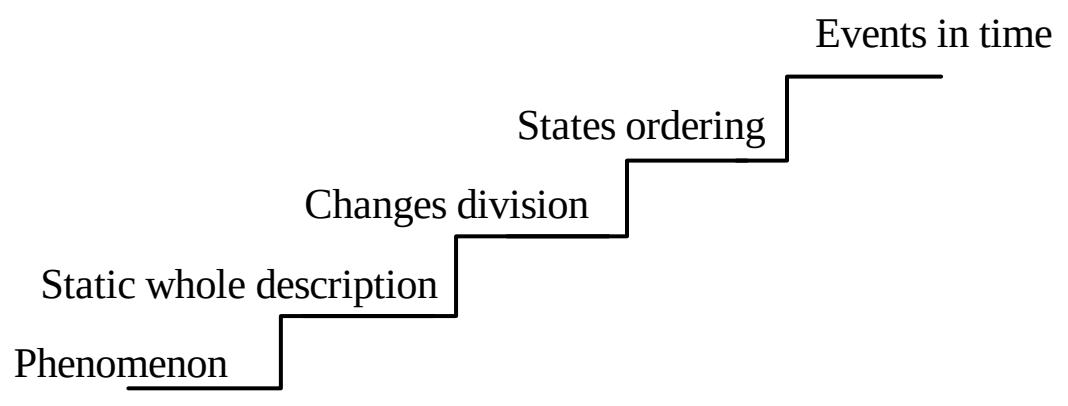

**Fig. 12. Progression of the stages from staticity to dynamism.**

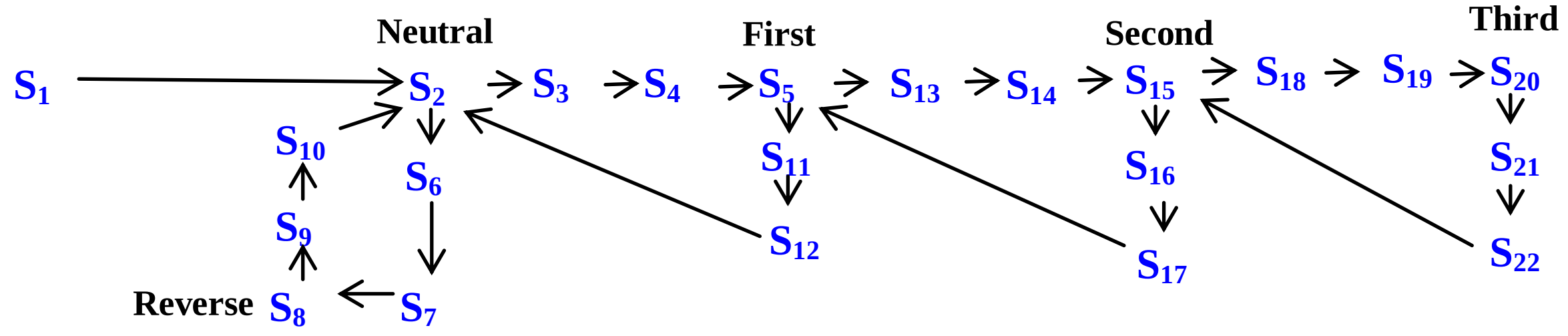

**Fig. 11. The chronology of the transmission system's static states.**

The chronology of states that specifies "before," "after," and "simultaneously" (Fig. 10) does not designate a dynamic (temporal) change when we go from one state to another. However, this (static) "subtransformation" refers to the whole (e.g., S) stimulating its parts to appropriate their roles from each other (e.g., the transmission in the "third-gear position" state shifting to the "second-gear position" state). The static states can be converted to events when the time is brought into the static picture of the states' chronology (Fig. 13). Events, not states, are the genuine conveyers of behavior. Thus, Fig. 13 shows the system's behavior in terms of these events, described as follows.

Event 1: ($E_1$): Starting the car
Event 2: ($E_2$): Neutral is ready
Event 3: ($E_3$): Shifting from neutral to first gear
Event 4: ($E_4$): Shifting from neutral to first
Event 5: ($E_5$): Driving in first (car accelerates)
Event 6: ($E_6$): Selecting from neutral to reverse
Event 7: ($E_7$): Shifting from neutral to reverse
Event 8: ($E_8$): Driving in reverse
Event 9: ($E_9$): Selecting from reverse to neutral
Event 10: ($E_{10}$): Shifting from reverse to neutral
Event 11: ($E_{11}$): Selecting from first to neutral
Event 12: ($E_{12}$): Shifting from first to neutral
Event 13: ($E_{13}$): Selecting from first to second
Event 14: ($E_{14}$): Shifting from first to second
Event 15: ($E_{15}$): Driving in second
Event 16: ($E_{16}$): Selecting from second to first
Event 17: ($E_{17}$): Shifting from second to first
Event 18: ($E_{18}$): Selecting from second to third
Event 19: ($E_{19}$): Moving from second to third
Event 20: ($E_{20}$): Driving in third
Event 21: ($E_{21}$): Selecting from third to second
Event 22: ($E_{22}$): Shifting from third to second

Additional events can be added, such as waiting, interruptions, or warnings. Some of the events may be regionless events that can be interwoven with the events that originate from states.

## 7. Conclusion

This paper contributes to establishing a broad ontological foundation for the transformation from static modeling to specifying a system's behavior. Such a distinction of modeling phases involves replacing the static descriptive whole by the organization of its parts (e.g., by neighborhood). This transformation starts with dividing the static description into static changes. Static changes lead to a conception of static states (with more fine-tuned meaningfulness than those of FSM states) that contrasts with the sudden appearance of FSM states when developing the FSM model. The static states are further synthesized (in terms of realistic practicality) as events assembled by the establishment of a common temporality.

The significance of this discussion is its role in clarifying the notion of a state and its relationship to systems' behavior. Further research will connect this analysis with the classical philosophical notion of change and behavior.

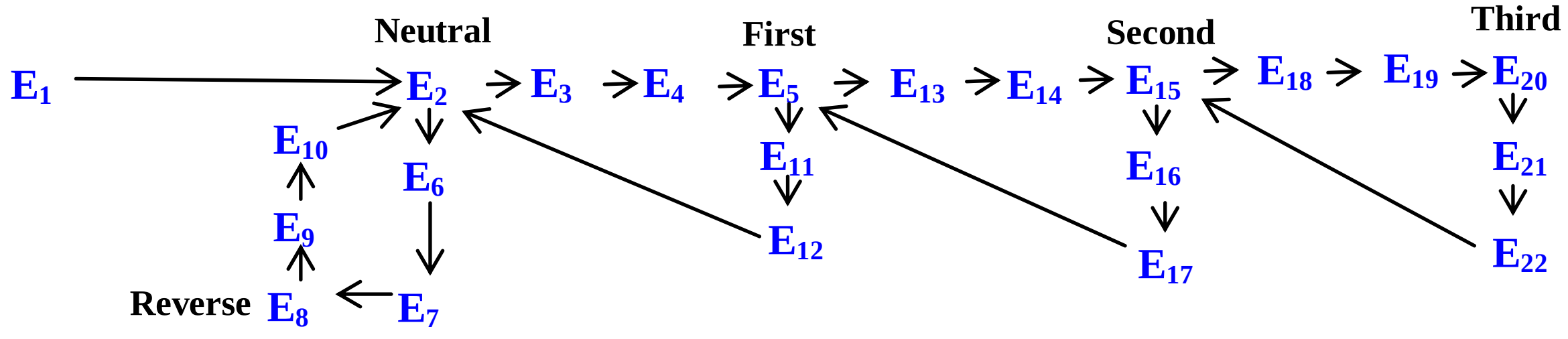

**Fig. 13. The chronology of static states of the transmission system.**